\documentclass[conference]{IEEEtran}
\ifCLASSINFOpdf
\usepackage[pdftex]{graphicx}
\else
\fi
\usepackage{amsmath}
\begin{document}

\title{Classical Tracking for Quantum Trajectories}

\author{\IEEEauthorblockN{Jason F Ralph\IEEEauthorrefmark{1},
Simon Maskell\IEEEauthorrefmark{1},
Michael Ransom\IEEEauthorrefmark{1}, 
Hendrik Ulbricht\IEEEauthorrefmark{2}}
\IEEEauthorblockA{\IEEEauthorrefmark{1}Department of Electrical Engineering and Electronics, 
University of Liverpool,  Brownlow Hill, Liverpool, L69 3GJ, UK.\\
Email: \{jfralph, smaskell, M.J.Ransom\}@liverpool.ac.uk}
\IEEEauthorblockA{\IEEEauthorrefmark{2}Department of Physics and Astronomy, 
University of Southampton, University Road, Southampton, SO17 1BJ, UK\\
Email: H.Ulbricht@soton.ac.uk}}

\maketitle

\begin{abstract}
Quantum state estimation, based on the numerical integration of stochastic master equations (SMEs), provides estimates for the evolution of quantum systems subject to continuous weak measurements. The approach is similar to classical state estimation methods in that the `quantum trajectories' produced by solving the SME are conditioned on continuous classical measurement signals. In this paper, we explore the use of classical state estimation for a candidate quantum system, one based on an experimentally realisable system: a material object undergoing continuous feedback cooling in an optical trap. In particular, we demonstrate that classical tracking methods based on particle filters can be used to track quantum states, and are particularly useful for higher temperature regimes where quantum state estimation becomes computationally demanding. 
\end{abstract}


\IEEEpeerreviewmaketitle

\section{Introduction}\label{intro}

The development of quantum analogues for classical state estimation originated with Belavkin in the early 1980s~\cite{Bel1999}, and appeared independently in the late 1980s and early 1990s as stochastic Schr\"{o}dinger equations or `quantum trajectory' methods. These methods go beyond the standard description of measurement in quantum mechanics, which tend to present measurements as projection operations. That is, the act of measurement alters the state of an otherwise isolated system in a fundamental way, localising the state in one variable at the expense of others. By contrast, quantum state estimation methods are used to describe the continuous evolution of open quantum systems; a quantum system coupled to a dissipative environment~\cite{Wis2010,Jac2014}. The action of the dissipative environment is to encode aspects of the quantum behaviour, such that the combined system can be subjected to repeated projective measurements that record the continuous quantum evolution whilst minimising the disturbance of the quantum state due to the back action of the measurements. The probabilistic nature of quantum measurements means that the back action perturbs the quantum state and acts as a source of noise which is correlated with the measurement signal. The resulting stochastic master equation (SME) contains: the familiar evolution Schr\"{o}dinger term, the effect of dissipation, and a stochastic innovation term. These stochastic models have been studied extensively as part of the theory of quantum feedback control~\cite{Wis2010,Jac2014}, but their importance has grown since the demonstration of the reconstruction of quantum trajectories in experimental superconducting systems~\cite{Mur2013,Web2014,Six2015,Cam2016}. Quantum state estimation methods have been studied extensively in the quantum physics and quantum optical communities. Bayesian methods have been developed for quantum systems~\cite{Kor99}, but are rarely studied using modern, computationally efficient tracking methods. 

In this paper, we examine the use of classical tracking methods to estimate the state of a nonlinear quantum oscillator that is subject to feedback cooling. Whilst quantum state estimation can be used for such systems, the numerical solution of the SME becomes computationally demanding when the kinetic temperature of the oscillator is above the lowest few energy states. Solving the SME requires the numerical integration of a state that is defined by a complex-valued Hermitian matrix (the density matrix, $\rho = \rho^{\dagger}$) and physical quantities are represented by operators -- matrices that are (in general) non-commutative. Added to this, the density matrix must retain a unit trace and be positive semi-definite to represent a physical state, and -- when dealing with thermal environments -- several noise sources may be required, leading to non-commutativity in a classical stochastic sense rather than a quantum one. 

The system considered in this paper can be mapped to a practical experimental system: a levitated optomechanical system, consisting of a microscopic bead of fused silica that is held in an optical trap by a focussed laser and in contact with an ultra low pressure buffer gas~\cite{Ash1970,Man1998,Gan1999,Vul2000,Mud2016,Set2018}. The aim for such experiments is often to test the limits of current physical theories, such as modifications to gravity at the small scale~\cite{Ger2010,Arv2013} or quantum mechanics at scales larger than single atoms or molecules~\cite{Bas2003,Bas2013,Ber2015}. Even though the silica bead is microscopic, it is large enough to contain thousands or tens of thousands of atoms, so it can be considered to be large when compared to more familiar quantum systems. Experiments with these systems is of great interest currently because quantum mechanical behaviour has recently been demonstrated in these systems~\cite{Del2020} but reaching the temperatures required for such experiments is still very challenging. The aims of the current paper are to demonstrate the utility of classical tracking methods based on conventional particle filters~\cite{Gor1993,Dou2001,Aru2002} in reconstructing trajectories and probability density functions from classical measurement records arising from the evolution of a quantum system, and to assist in the development of quantum control methods for the exploration of fundamental physics. 

We begin in \ref{qse} by providing an overview of quantum trajectory methods and the use of the SME to generate a quantum state conditioned on a classical measurement record. In this, we attempt to emphasise the links to classical state estimation. In \ref{model}, we introduce a model that can be used to describe the optomechanical system of interest. Section \ref{results} presents some example results, including a comparison of the quantum and classical distributions: including a qualitative comparison in \ref{Wigner} and quantitative comparison using the Kullback-Leibler divergence for a range of parameter values in \ref{KLdiv}. We then summarise the work in \ref{conclusions}.

\section{Quantum State Estimation}\label{qse}

The most familiar way to describe the state of a quantum system is through the construction of a wave function, $\psi(x)$, whose evolution is described by the Schr\"{o}dinger equation. The wave function is complex valued and its magnitude squared $|\psi(x)|^2$ provides a probability density function (pdf) for the result of direct measurements of $x$. Many of the conceptual difficulties in the interpretation of quantum mechanics stem from questions about the objective nature of the wave function, and the relationship between the Schr\"{o}dinger equation and the action of a measurement to project the wave function into a state corresponding to the measured value. For position measurements, this projective action localises the wave function instantaneously around the measured position. 

An alternative, and more general, description of a quantum state is the density matrix, $\rho$, which can be used to represent probability distributions over a set of possible wave functions, 
\begin{eqnarray}\label{densityOp}
\rho&=&\sum_{i} P_i |\psi_i\rangle\langle \psi_i|
\end{eqnarray}
where the $P_i$'s are probabilities associated with a set of possible wave functions, the wave functions are written using the {\it bra-ket} notation~\cite{Wis2010,Jac2014}, and the product in the summation is an outer product of the {\it ket} and the {\it bra}. For practical problems, the density matrix is formed using basis states and the quantum state is represented as an expansion of these basis states. Depending on the properties of the system, the number of basis states can be very large if a suitable basis cannot be found. It is not uncommon for a density matrix to require hundreds of basis states. 

A wave function solution to the Schr\"{o}dinger equation is said to be a `pure' state. By contrast, a density matrix can be used to represent `mixed' states with purities less than one, where the purity is defined as $\mathrm{Tr}[\rho^2] $, where $\mathrm{Tr}[.]$ is the trace operator. In mixed states, the quantum state is uncertain in a classical probabilistic sense as well as a quantum sense. 

This lack of certainty is familiar in classical state estimation. The state of a classical system is only known through the measurements that have been made, and the information that they provide. The estimated classical state represents all of the knowledge that has been accumulated from a series of measurements. The big difference between classical and quantum systems is that the effect of the measurement on a quantum state cannot be reduced to zero. Even though all practical measurements will have an effect on the evolution of a classical system, it is possible -- in principle at least -- to reduce the effect of the measurement back action to zero. In quantum systems, this is not possible, even in principle. In addition, a wave function or a density matrix cannot be measured directly. They can only be inferred from a sequence of measurements of other physical quantities, such as position or momentum -- and, even so, different combinations of wave functions that generate the same density matrix cannot be distinguished physically. In the standard approach to quantum measurement, a measurement alters the quantum state to reflect the value that has been measured. This projective property of measurement requires that a quantum state must be re-prepared and re-measured a large number of times to build up statistics from which the wave function or density matrix can be reconstructed: a process known as {\it quantum tomography}.

Rather than inferring the quantum state from a series of repeated projective measurements, a quantum trajectory is constructed from a series of continuous weak measurements of the system. To do this, the quantum system is coupled to an environment and is said to be {\it open}. An open system experiences dissipation and other perturbations due to environmental noise. This allows the back action of the measurement on the quantum evolution to be reduced by considering the coupled system (system of interest and environment) as a single system that is projected after each measurement/time increment. The measurements on the coupled system contain aspects of the quantum evolution without fully projecting the quantum state of the sub-system of interest. Assuming that the environment is sufficiently dissipative that any quantum correlations between the system of interest and the environment disappear quickly\footnote{Quick, relative to the time increment. This is a Markov approximation, but for most practical systems it is a reasonable simplification.}, the environment can be averaged out to provide a continuous weak measurement record that carries information about the quantum state of the system of interest and its evolution~\cite{Wis2010,Jac2014}; albeit an evolution perturbed by the stochastic effect of the environment. 

The equation that describes the evolution of the density matrix of an individual open quantum system subject to a continuous weak measurement is the stochastic master equation~\cite{Bel1999,Wis2010,Jac2014}. The average evolution of an ensemble of open quantum systems is described by the master equation, and focussing on one particular member of an ensemble (one realisation from an ensemble) leads to the SME. This is often referred to as an `unravelling' of the master equation, and there are an infinite number of such unravellings, which correspond to different ways in which the measurement can be taken~\cite{Wis2005}. For simplicity, we will focus on one particular unravelling, which generates an SME given by~\cite{Wis2010,Jac2014},
\begin{eqnarray}\label{sme1}
d\rho&=&- i \left[\hat{H},\rho\right]dt \nonumber\\
&&+\left( \hat{L}\rho \hat{L}^{\dagger} -\frac{1}{2}\left(\hat{L}^{\dagger} \hat{L} \rho 
+ \rho \hat{L}^{\dagger}\hat{L} \right)\right)dt   \nonumber\\
&&+ \sqrt{\eta}\left(\hat{L}\rho+\rho \hat{L}^{\dagger}-\mathrm{Tr}[\hat{L}\rho+\rho \hat{L}^{\dagger}] \right)dW \nonumber\\
&&+\sum_{j=1}^{m} \left\{ \hat{V}_{j} \rho\hat{V}^{\dagger}_{j} -\frac{1}{2}\left(\hat{V}^{\dagger}_{j} \hat{V}_{j} \rho 
+ \rho \hat{V}^{\dagger}_{j} \hat{V}_{j} \right)\right\}dt  
\end{eqnarray}
where $\hat{H}$ is the Hamiltonian of the quantum system, $dt$ is an infinitesimal time increment, and we set the reduced Planck's constant $\hbar = 1$ for convenience. The effect of the environment is represented by two types of quantum operator: one operator for the measurement $\hat{L}$, and a set of operators that do not generate a measurement record and give rise to environmental dissipation $\hat{V}_j$  $(j = 1\dots m)$. The measurement and environment operators can be Hermitian, $\hat{L} =\hat{L}^{\dagger}$, but this is not necessary. However, physically observable quantities are always represented by Hermitian operators. In this unravelling, $dW$ is taken to be a real Wiener increment such that $E(dW)=0$ and $dW^2  = dt$ for simplicity. Other choices lead to alternative unravellings and different trajectory models~\cite{Wis2005}.

In the ideal case, the measurement record would be 100\% efficient, with all of the available information being reflected in the measurement record. Unfortunately, real measurements are always diluted by additional noise sources, which can be parameterised using an efficiency parameter, so that $\hat{L}$ has an efficiency $\eta$. Specifically, $\eta$ is the fraction of the total noise power due to the quantum measurement compared to the power contained in the other additional noise sources. For the SME given by (\ref{sme1}), the measurement record for the operator $\hat{L}$ during a time step $t\rightarrow t+dt$ is given by, 
$$
y(t+dt)-y(t)=dy(t)= \sqrt{\eta}\mathrm{Tr}[\hat{L}\rho_c+\rho_c \hat{L}^{\dagger}] dt+dW
$$
In this paper, we use two SMEs: one to simulate a classical measurement record, and one to take the record and to estimate the quantum state conditioned on it. This two stage process allows us to generate a suitable measurement record with the first SME, in the absence of an experimental system, and to provide a reconstructed quantum state estimate to compare against the output of the classical tracker. 
The second SME takes the measurement record and integrates an equivalent SME to generate a conditional density matrix $\rho_c$ representing the estimated quantum states given the measurement record produced by the first SME. In classical tracking terms, the first SME provides the ground truth for the quantum evolution, and the second provides the optimal quantum state estimation; it is optimal but computationally challenging.  

The second SME starts with an initially unknown quantum state, represented by a completely mixed density matrix (analogous to a uniform or uninformative prior in classical state estimation), and takes the measurement record $dy(t)$ and forms a stochastic increment,
$$
d\tilde{W}= dy(t) - \sqrt{\eta}\mathrm{Tr}[\hat{L}\rho_c+\rho_c \hat{L}^{\dagger}] dt
$$
This is the difference between an actual measurement $dy(t)$ and a measurement that is estimated from the current conditional density matrix (the trace of an operator with a density matrix is the expectation value of that operator for that state). As such, $d\tilde{W}$ is equivalent to an innovation that might be used in classical state estimation. 

Given this innovation, the conditional density matrix can be updated by integrating the same SME as the first (\ref{sme1}) with $\rho_c$ replacing $\rho$ and $d\tilde{W}(t)$ replacing $dW$. Integrating the SME and conditioning the state on the measurement record gradually purifies the mixed state and the density matrix starts to track the dynamics of the quantum system of interest. However, for a quantum system perturbed by external noise, the density matrix will remain slightly mixed. The degree of impurity\footnote{{\it Impurity} is defined as one minus purity, $1-\mathrm{Tr}[\rho^2]$} that remains is an indication of the residual classical uncertainty in the density matrix and the classical probabilities used in its construction. In this case, first and second SME provide, after an initial transient period, the same quantum trajectories but, for more complex situations, the approach can be generalised to examples where there are model mismatches between the first simulation SME and the second estimation SME and be used to estimate system parameters~\cite{Ral2017}.

To model the coupling of the system of interest to the ambient environment (in our case, a buffer gas at ultra low pressure and finite temperature), we include a thermal environment which is modelled with two unmeasured environmental operators, 
\begin{eqnarray}\label{thermalOps}
\hat{V}_1 = \, \sqrt{\frac{(\bar{n}+1)\omega}{Q}} \, \hat{a} ,  \;\;\;\;  \hat{V}_2  = \, \sqrt{\frac{\bar{n}\omega}{Q}} \, \hat{a}^{\dagger} . 
\end{eqnarray}
where $\hat{a}^{\dagger}$ and $\hat{a}$ are the raising and lowering operators for the quantum oscillator respectively, $\omega$ is the angular frequency of the oscillator, $Q$ is its quality factor, and \mbox{$\bar{n} = [\exp(\hbar\omega/k_{B} T) - 1]^{-1}$} where $k_{B}$ is Boltzmann's constant and $T$ is the temperature of the environment. The first operator represents thermally induced decay from one energy level to the one below, and the second provides thermal excitations from one energy level up to the next. Combining the two environmental operators provides a quantum model for conventional classical damping due to a thermal environment~\cite{Spi1993}.  

\section{Levitated Optomechanical Bead in a Trap}\label{model}

The system studied in this paper is based on a model for an experimental system consisting of a microscopic, fused silica bead that is held in an optical trap, as described in~\cite{Ash1970,Man1998,Gan1999,Vul2000,Mud2016,Set2018}. The bead is typically around $10^{-18}$kg, which is small but still contains thousands or tens of thousands of atoms and, for a quantum system, can be considered to be large. A laser beam is focussed onto a small parabolic mirror so that a trapping potential forms at the focal point of the mirror. The bead is loaded into the trap, inside a vacuum chamber filled with a low pressure buffer gas. The scattered laser light is monitored and the motion of the bead is encoded in the phase of this scattered light. Measuring the phase in this way provides a continuous weak measurement of the position of the bead. The effect of measurement on a quantum bead is to localise it in the position basis, but the fact that it is a weak continuous measurement of the bead and its environment means that the measurement has an effect on its motion but this effect shows up only as a back action noise term in the SME. This is one principal difference between the quantum model (SME) and an equivalent classical model, where there will be an additional noise term due to measurement but, in the quantum case, this measurement noise cannot be removed and it is correlated with the measurement signal. When the effect of the measurement is very small compared to the other noise sources (such as thermal noise from the environment) the classical and quantum models should be indistinguishable. 

The trap potential is nonlinear, containing a quadratic term and a weaker quartic term, and can be modulated by changing the input laser. This modulation is used to cool the motion of the bead, stiffening the trap when the bead is towards the edges of the trap and softening it when the bead is near the centre of the trap. The excess kinetic energy is removed by collisions with the buffer gas, and -- as the system cools down -- the pressure of the buffer gas is gradually reduced to prevent the collisions heating the bead up again. This process can cool the motion of the bead to very low temperatures. Of course, the motion of a bead within a real trap has three translational and three rotational degrees of freedom~\cite{Rom2011,Sti2016,Ral2016}. However, at low temperatures, the different degrees of freedom can be isolated and can be studied independently. The system has been studied extensively and a related system, albeit with a different cooling mechanism, has recently been demonstrated to behave quantum mechanically in the very low temperature limit~\cite{Del2020}. To simplify the analysis, we use a dimensionless form for the dynamical model, but this can be mapped back onto experimental parameters relatively easily~\cite{Gie2015}. Firstly, the quantum Hamiltonian of a bead in a one dimensional nonlinear trap subject to feedback cooling is given by,
\begin{equation}\label{DuffHam2}
\hat{H} = \frac{1}{2}\hat{p}^2+(1+\alpha\sin(2\phi))(\frac{1}{2}\hat{x}^2+\frac{\gamma}{4}\hat{x}^4)+\frac{\Gamma}{4}(\hat{x}\hat{p}+\hat{p}\hat{x})
\end{equation}
where we have set both the mass of the bead and the linear oscillator frequency $\omega$ to be one. The damping coefficient is then $\Gamma = 1/Q$. The term $(1+\alpha\sin(2\phi))$ provides the feedback cooling of the bead in the trap, where the phase is given by the expectation values of the dynamical quantities,
$$
\phi = \tan^{-1}{(\mathrm{Tr}[\hat{p}\rho]/\mathrm{Tr}[\hat{x}\rho])}
$$
Typically the feedback coefficient in experiments is around a few percent, $\alpha\simeq 0.02-0.04$. The final term in the Hamiltonian is included for numerical reasons. It has no effect on the equivalent classical equations and it only serves to mix the effect of dissipation equally between the position and momentum co-ordinates~\cite{Bru1996}. 

Using a continuous weak measurement of position, corresponding to a measurement operator $\hat{L}= \sqrt{8k}\hat{x}$, with a measurement strength $k$ and a measurement efficiency $\eta$, the SME reduces to,
\begin{eqnarray}\label{sme2}
 d\rho & = & -i [\hat{H}, \rho] dt - k [\hat{x},[\hat{x},\rho]] dt  \nonumber \\ 
                 &&+ \sqrt{2k\eta}(\hat{x} \rho + \rho \hat{x} - 2 \mbox{Tr}[\hat{x}\rho] \rho ) dW \nonumber \\ 
                 && + \sum_{j= 1,2} \left\{\hat{V}_{j} \rho \hat{V}^{\dagger}_{j} -\frac{1}{2}\left(\hat{V}^{\dagger}_{j} \hat{V}_{j} \rho + \rho\hat{V}^{\dagger}_{j} V_{j} \right)\right\}dt \nonumber \\ 
\end{eqnarray} 
which generates a measurement record,
$$
dy(t) = \sqrt{8k\eta} \mathrm{Tr}[\hat{x}\rho] dt + dW(t)
$$
For the numerical integration of the SME, we use Rouchon's method  to ensure the positivity of the density matrix~\cite{Ami2011,Rou2015}. For the examples shown in this paper, the quantum SME has been solved using 100-120 linear oscillator basis states to construct the density matrix, which is sufficient for $k_B T\le 2$ (or $\bar{n}\simeq 1.5$) but the number of states required increases rapidly as the temperature is increased. 

The equivalent classical stochastic differential equation (SDE) for this system is given by~\cite{Hop2003},
\begin{eqnarray}\label{sde1}
dp &= &(1+\alpha\sin(2\phi))(-x - \gamma x^3)dt-\Gamma p dt \nonumber \\
       &&+ \sqrt{2k}dY+\sqrt{2 \Gamma k_B T}dU \nonumber \\ 
dx  &=& p dt 
\end{eqnarray}
where the classical measurement record is $d\tilde{y}(t) = \sqrt{8k\eta} x dt+dZ(t)$ and the phase in the feedback term is $\phi = \tan^{-1}(p/x)$. The three noise sources in the SDE are independent real Wiener increments,  with $E(dZ)=E(dY)=E(dU)=0$, $dZ^2=dY^2=dU^2=dt$ and $dZdU=dZdY=dYdU=0$. Note that the classical SDE includes a noise term due to the effect of the measurement but this is uncorrelated with the measurement noise because there is no coherent back action from the measurement on the state from a classical measurement as there would be for a quantum system~\cite{Hop2003}.

\section{Quantum and Classical Trajectories}\label{results}

In classical state estimation, the properties of the bead are specified in terms of its state vector, generally including dynamical variables such as position and velocity. The probability density function for the state vector can be modelled, either using statistical methods (such as the mean and covariance, as in the Kalman filter and its nonlinear variants~\cite{Aru2002}) or using the weights attached to particles in particle filters or other importance sampling methods~\cite{Gor1993,Dou2001,Aru2002}. 

In quantum mechanics, it is more common to use position and momentum as the dynamical variables, but position and momentum are incompatible  in quantum mechanics. The operators for position and momentum do not commute and they cannot be simultaneously defined, which is the origin of the famous Heisenberg uncertainty relations~\cite{Wis2010,Jac2014}. As noted above, to define a pdf for one or the other is possible by calculating the squared amplitude of the complex wave function, but it is not possible to define a joint pdf for both position and momentum because of the incompatibility of the measurement operators. The most common alternative approach, which does allow a quantum distribution to be visualised in state space (often called {\it phase space} in physics) is the Wigner quasi-probability distribution~\cite{Til2016}, which is defined for one degree of freedom with a density matrix $\rho$ by,
\begin{equation}
W(x,p) = \left(\frac{1}{2\pi}\right)\int_{-\infty}^{\infty}dz\langle x-\frac{z}{2}\left|\rho\right|x+\frac{z}{2}\rangle e^{-xp}
\end{equation}
The Wigner function has most of the properties of the probability density function -- its integral over phase space is one and expectations can be calculated by integrating over $x$ or $p$ and weighting by the quasi-probability, for example -- but it also allows negative values~\cite{Sze2005}. Negative values in the Wigner distribution are associated with manifestations of the quantum nature of the system, either interference arising from superposition of distinguishable states or entanglement\footnote{Entanglement is a characteristic of quantum systems where the correlations between systems are `stronger' than any classical correlations, and it is often associated with non-locality and the performance increase available from quantum computing.}. However, for most values of $x$ and $p$, the Wigner function is positive and it closely matches the properties of a classical pdf. We would expect that a classical pdf would be able to represent most of the properties of the Wigner function, but not the negative quasi-probability values associated with them. The adaptation of classical methods to represent negative pseudo-probabilities in a way consistent with quantum evolution is an open question. As noted above, quantum Bayesian methods have been proposed~\cite{Kor99}, efficient numerical integration methods developed~\cite{Ami2011,Rou2015}, and these have been tested in experiments~\cite{Mur2013,Web2014,Six2015}. However, this work has not benefited significantly from recent advances in classical state estimation and associated computational methods. As a result, we do not aim to solve the entire problem in this paper. 

We note that -- for higher temperatures, above the lowest few energy states of a quantum system when it would be expected to be more classical than quantum -- the agreement between the classical and quantum models should be good (i.e. the classical behaviour is recovered in this limit), and we find that the agreement is also good in the lower temperature regimes, at least for regions of the pdf where a classical particle filter can provide accurate representations of the positive Wigner distributions. As a minimum, this would assist in the development of classical feedback controls for quantum systems even in the low temperature regime.

\begin{figure}[htbp]
       \centering
		\includegraphics[width=0.85\hsize]{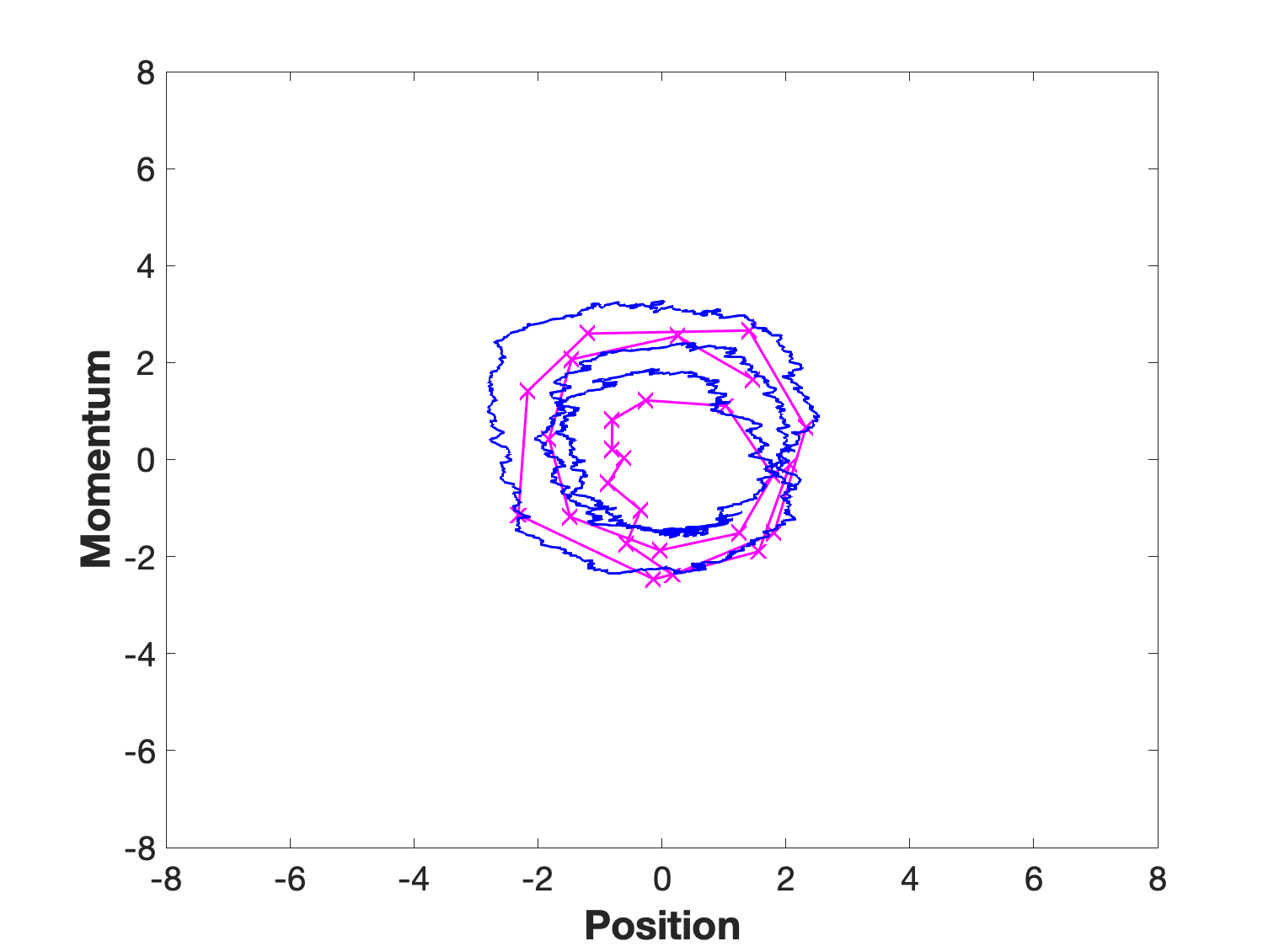} 
			\caption{\label{trajectory} Example section of a quantum trajectory (blue) and the classical state estimates (magenta), the classical state estimates are calculated at each time step but are only plotted every 100 time increments for clarity, parameter values are given in the text.}
\end{figure}
We start by providing an example of the quantum and classical trajectories. For the classical trajectory, we have used a standard `bootstrap' particle filter~\cite{Gor1993,Dou2001,Aru2002} with standard importance sampling and the motion model given in equation (\ref{sde1}). The number of particles for the data shown is 1000 per particle filter and the threshold for resampling is $N_{thresh} = N_{eff}/2$, where the effective number, $N_{eff} = 1/(\sum_i (\tilde{w}^{(i)})^2)$ and the normalised particle weights are denoted by $w^{(i)}$. The weight update for the $i$'th particle after a time increment is given by,
\begin{equation}
\tilde{w}^{(i)}(t+dt) = \exp\left( -\frac{(y(t) - \sqrt{8k\eta} x^{(i)}dt)^2}{2 dt} \right)w^{(i)}(t)
\end{equation}
where $\tilde{w}^{(i)}(t)$ is the unnormalised weight after the update, $x^{(i)}$ is the particle position, and we have discretised the equations so that $dt \rightarrow \Delta t = 0.001$ (1000 steps per oscillation cycle of the oscillator, remembering that $\omega = 1$ in the scaled model). The choice of a bootstrap filter is pragmatic, in that it is straightforward to implement and to understand. As noted above, the application of modern tracking methods to quantum trajectories has not been widely studied, and even a relatively simple example does provide a good representation of the quantum evolution. Simple methods do not address issues around the estimation of the negative pseudo-probabilities and aspects associated with the `curse of dimensionality'~\cite{Dau2005,Sur2019}, which can inhibit the ability of particle filters to represent distributions in high dimensional settings. These issues are not dealt with in this paper, but we note the ability to extend the methods presented here using more sophisticated estimation methods.

Figure~\ref{trajectory} shows a section of one trajectory from an example generated using: $\eta = 1.0$, $k = 0.05$, $\gamma = 0.1$, $\Gamma = 0.125$, $k_B T = 2.0$, and $\alpha = 0.05$. The actual quantum trajectory (in blue) shows the system oscillating around the minimum of the trapping potential and subject to a large level of noise. The motion is driven by the noise and, over longer times, the system goes through periods when its oscillations are large (fully exploring the region of phase space shown in Figure~\ref{trajectory}), and other periods where it is close to the centre of the potential. Temperature, after all, is an average measure of the kinetic energy present in a system. In both extremes, however, the widths of the distributions (classical and quantum) are significant, similar to the examples shown in Figure~\ref{WQPF} below. 

\begin{figure}[htbp]
       \centering
		\includegraphics[width=0.85\hsize]{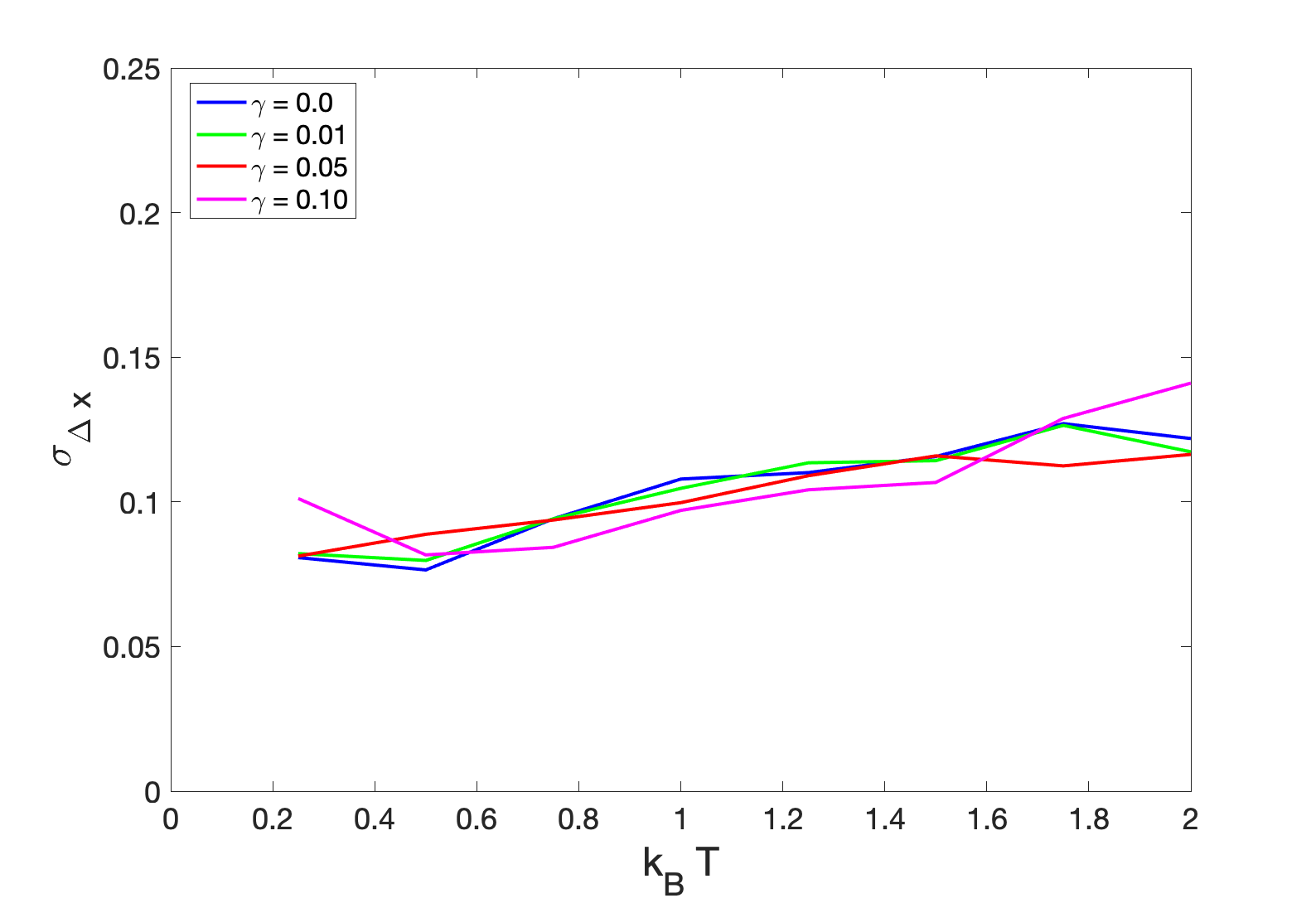} \\
		\includegraphics[width=0.85\hsize]{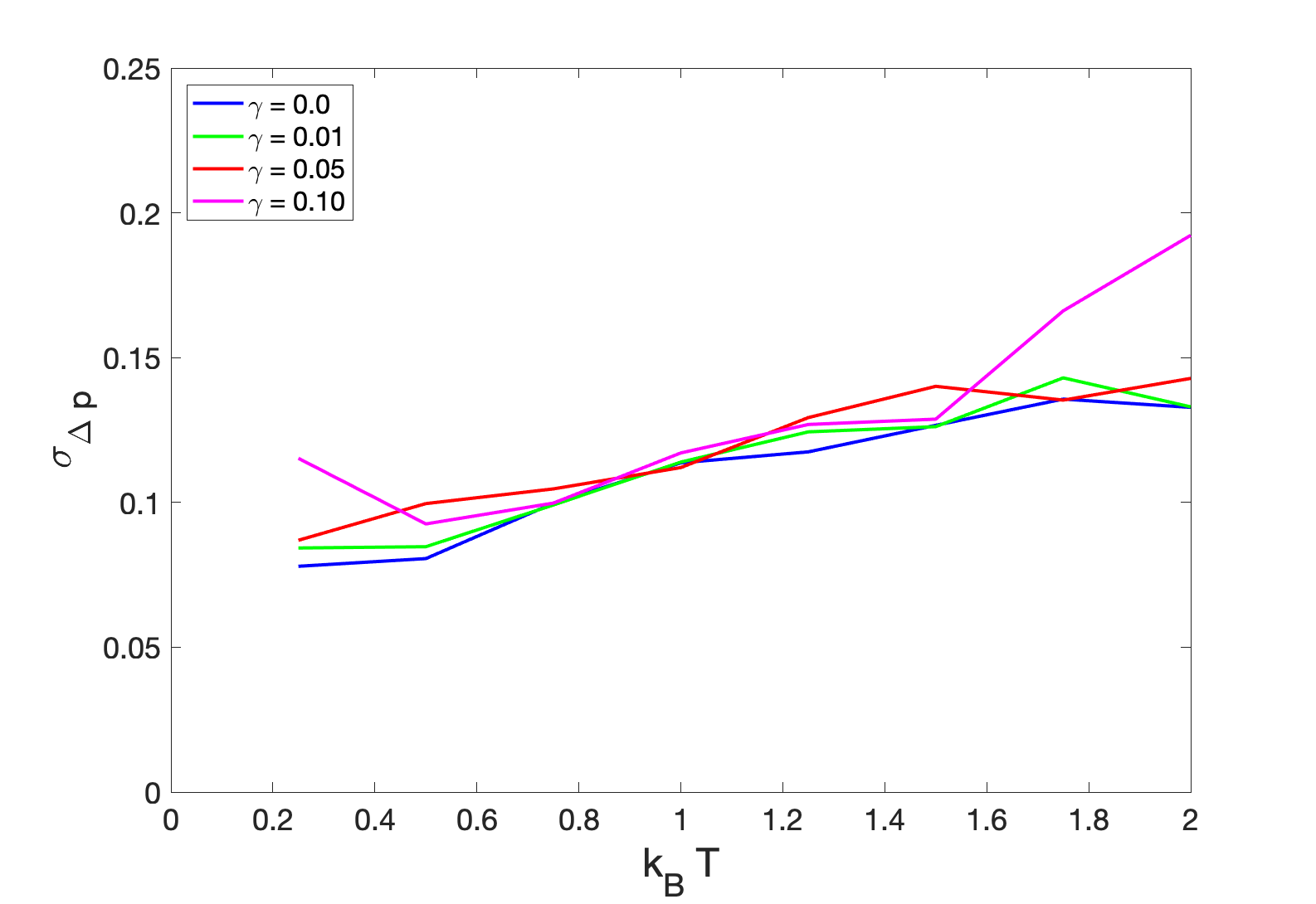} 
	\caption{\label{errors} Examples of standard deviations for errors between trajectories generated by the particle filter and the conditional SME for different values of the nonlinearity parameter $\gamma$, other parameter values are given in the text.}
\end{figure}
Figure~\ref{errors} shows the standard deviations of the average errors for between the mean position and mean momentum for the quantum and classical trajectories averaged over two hundred oscillator cycles with $\eta = 1.0$, $k = 0.05$, $\Gamma = 0.125$, and $\alpha = 0.05$ for a range of temperatures $k_B T = \{0.25, 0.5,...2.0\}$ and for four different values for the nonlinearity: $\gamma = \{0.0, 0.01, 0.05, 0.10\}$. The standard deviation for the errors does increase as the temperature rises, but this is to be expected because the level of noise in the system increases at higher temperatures. However, there are no large changes in the accuracy of the trajectories either at the higher temperatures or at the lower temperatures, where quantum effects would be expected to dominate. It is clear in both of the graphs in Figure~\ref{errors}, that there are no significant differences in the accuracy of the trajectory reconstruction for any of the values of the nonlinearity parameter. The higher values for $\gamma$ are large enough to generate highly nonlinear states, and similar systems can exhibit quite exotic behaviour including chaos~\cite{Bru1996}. The robustness of the trajectories is encouraging, because it demonstrates that the classical particle filters contain enough flexibility to follow quantum trajectories for a wide range of temperatures and nonlinearities.

\subsection{Results -- Wigner Quasi-Probability Function and the Classical Probability Density}\label{Wigner}

\begin{figure}[htbp]
       \centering
		\includegraphics[width=0.85\hsize]{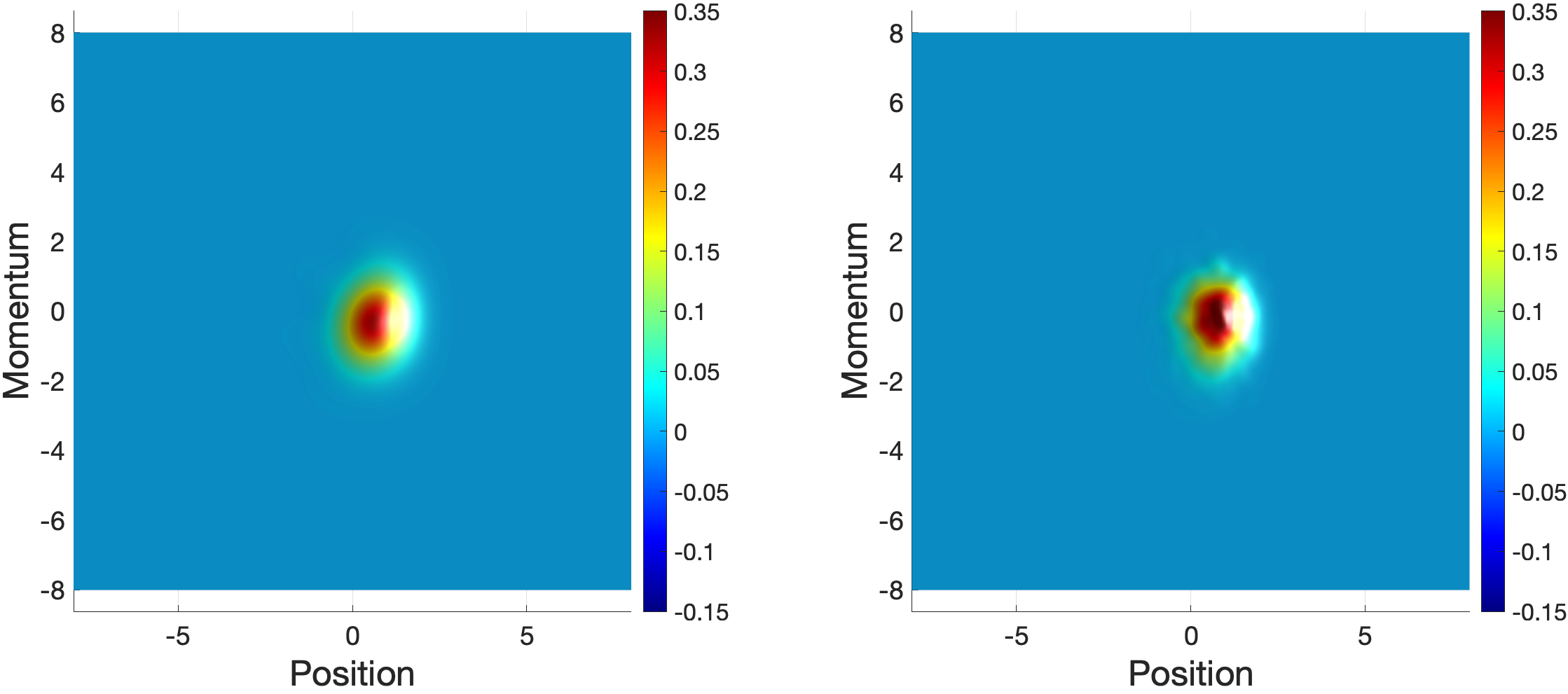} \\
		\includegraphics[width=0.85\hsize]{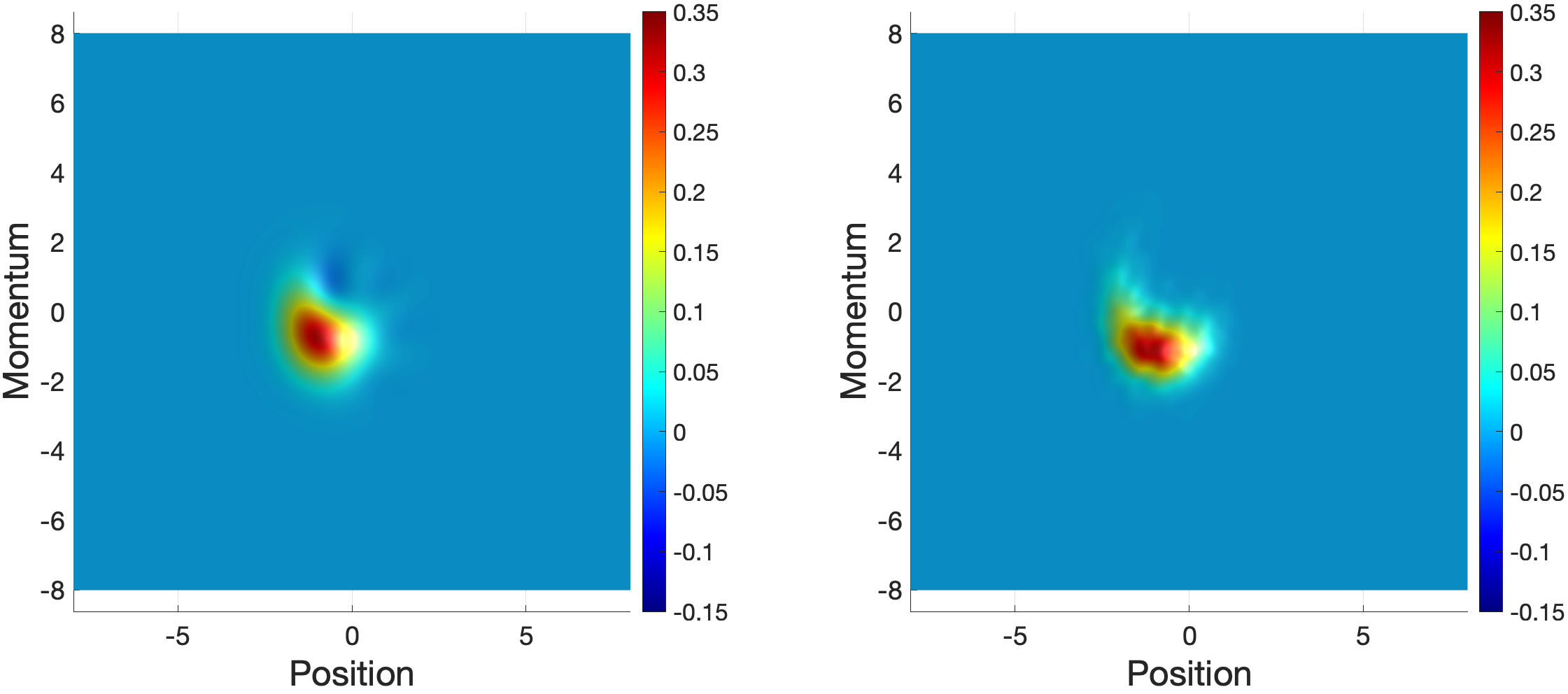} \\
		\includegraphics[width=0.85\hsize]{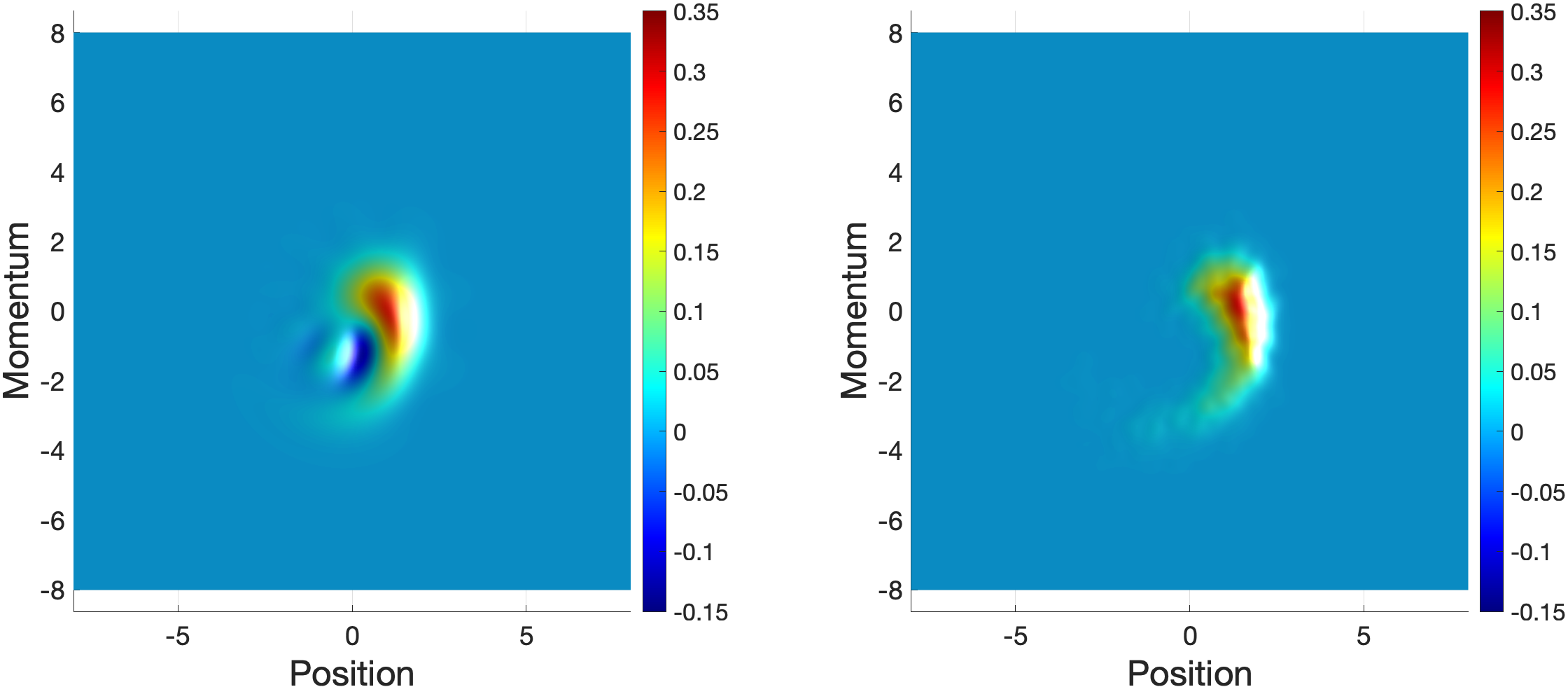}
	\caption{\label{WQPF} Three examples of Wigner quasi-probability distributions (left) and corresponding pdfs constructed from the particle weights (right) from one trajectory. The three examples show: classical-like state (top row); nonlinear quantum state with weak negative Wigner features (middle row); and nonlinear quantum state with strong negative Wigner features (bottom row). Nonlinearity $\gamma = 0.10$ with other parameter values  given in the text.}
\end{figure}

To compare the distributions arising from the quantum and the classical estimation methods, we calculate the Wigner quasi-probability distribution as described in \cite{Til2016} using a 201x201 grid covering the region of phase space shown in Figure~\ref{trajectory} and the classical pdf by summing the weights associated with the particles in each grid square. Examples of the Wigner functions and the associated classical pdfs are shown in Figure~\ref{WQPF} for $\eta = 1.0$, $k = 0.05$, $\gamma = 0.10$, $\Gamma = 0.025$, $k_B T = 2.0$, and $\alpha = 0.05$. The agreement for the three examples shown is good, except in regions where the Wigner distribution is negative. The classical particle filter cannot generate negative probabilities because it does not have the flexibility to be able to generate quantum mechanical behaviour. However, it does at least avoid placing particles in the regions where the Wigner function is negative. The examples shown are good examples but the stochastic nature of the particle filter does mean that there are some points along a trajectory where the agreement is not as good, but in these situations the particle filter does recover and reestablishes a match to the Wigner distribution within a few time steps. 

\subsection{Results -- Kullback–Leibler Divergence}\label{KLdiv}

Whilst it is encouraging that the particle filter is able to follow the classical aspects of the Wigner distribution qualitatively, it is important to have a quantitative metric for the similarity, and here we use the Kullback-Leibler (KL) divergence~\cite{Rol2012}, which is normally used to compare probability distributions. We use the discrete form of the KL divergence for the discrete grids defined for the Wigner function and the classical pdf described above,
\begin{eqnarray}
D[P^{(1)}({\bf x})||P^{(2)}({\bf x})] &=& \int d{\bf x} P^{(1)}({\bf x}) \ln\left(\frac{P^{(1)}({\bf x})}{P^{(2)}({\bf x})} \right) \nonumber \\
&\simeq &\sum_{i,j} P^{(1)}_{i,j} \ln\left(\frac{P^{(1)}_{i,j}}{P^{(2)}_{i,j}} \right) dx dp  \nonumber \\
\end{eqnarray}
where $P^{(1)}_{i,j} $ is the pdf constructed from the particle filter weights and $P^{(2)}_{i,j}$ contains the positive values from the Wigner quasi-probability distribution, both defined on 201x201 grids in phase space, and $dx$ and $dp$ are length of the grid elements in position and momentum, respectively. 

\begin{figure}[htbp]
       \centering
		\includegraphics[width=0.85\hsize]{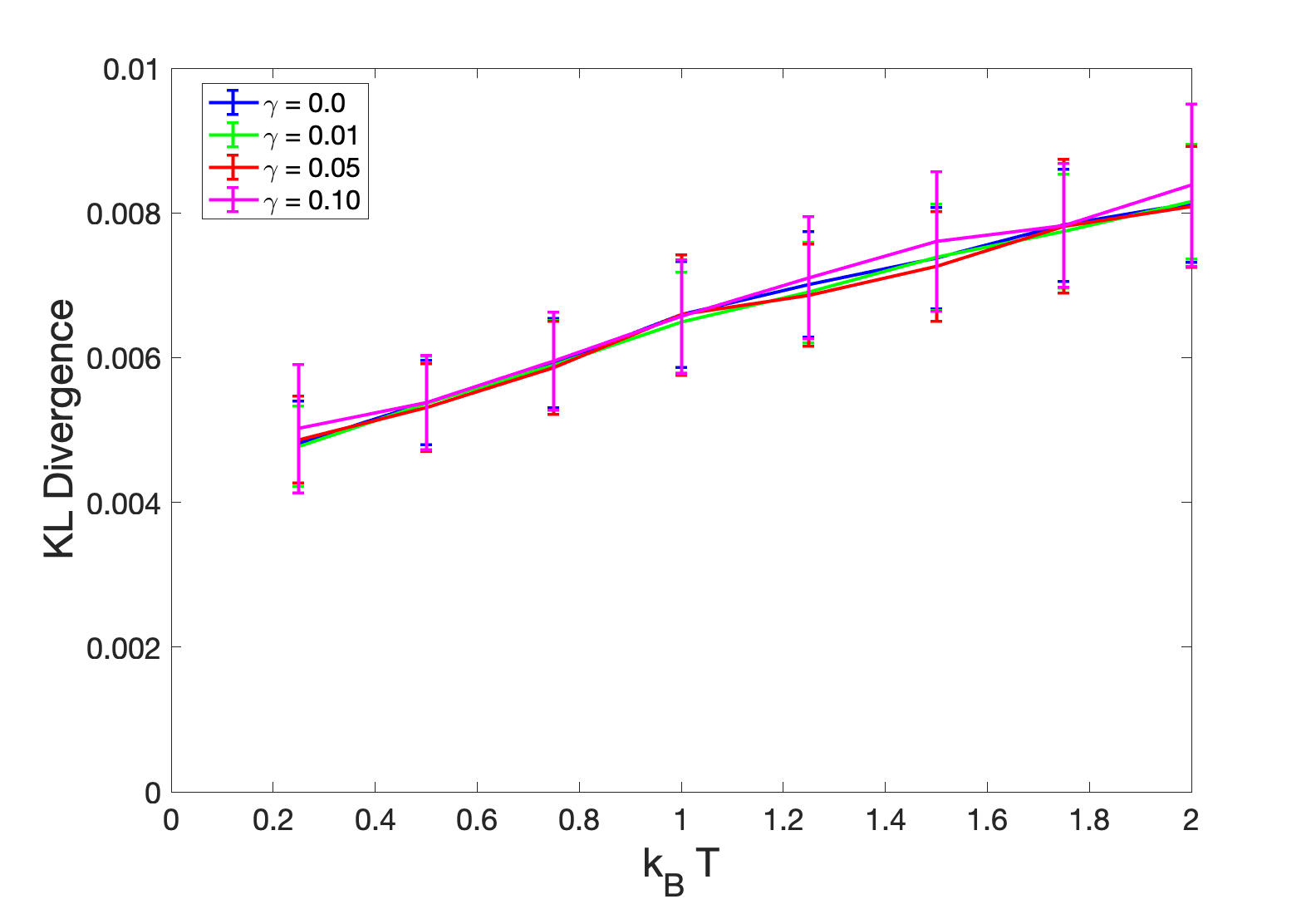} 
			\caption{\label{KL} Average values (with one standard deviation error bars) for KL divergences for the pdfs derived from the classical particle filter and the Wigner quasi-probability distributions for different values of the nonlinearity parameter $\gamma$, other parameter values are given in the text.}
\end{figure}
Figure~\ref{KL} shows the KL divergences and standard deviations (as error bars) for the trajectories used to generate Figure~\ref{errors}; $\eta = 1.0$, $k = 0.05$, $\Gamma = 0.125$, and $\alpha = 0.05$, $k_B T = \{0.25, 0.5,... 2.0\}$, and $\gamma = \{0.0, 0.01, 0.05, 0.10\}$. The KL divergences are remarkably consistent, showing no significant differences for the range of temperatures and nonlinearities used. These results demonstrate that the ability to generate accurate trajectories is robust to temperature, nonlinearity, and that distributions generated by the particle filter are also robust across these ranges. 

\begin{figure}[htbp]
       \centering
		\includegraphics[width=0.85\hsize]{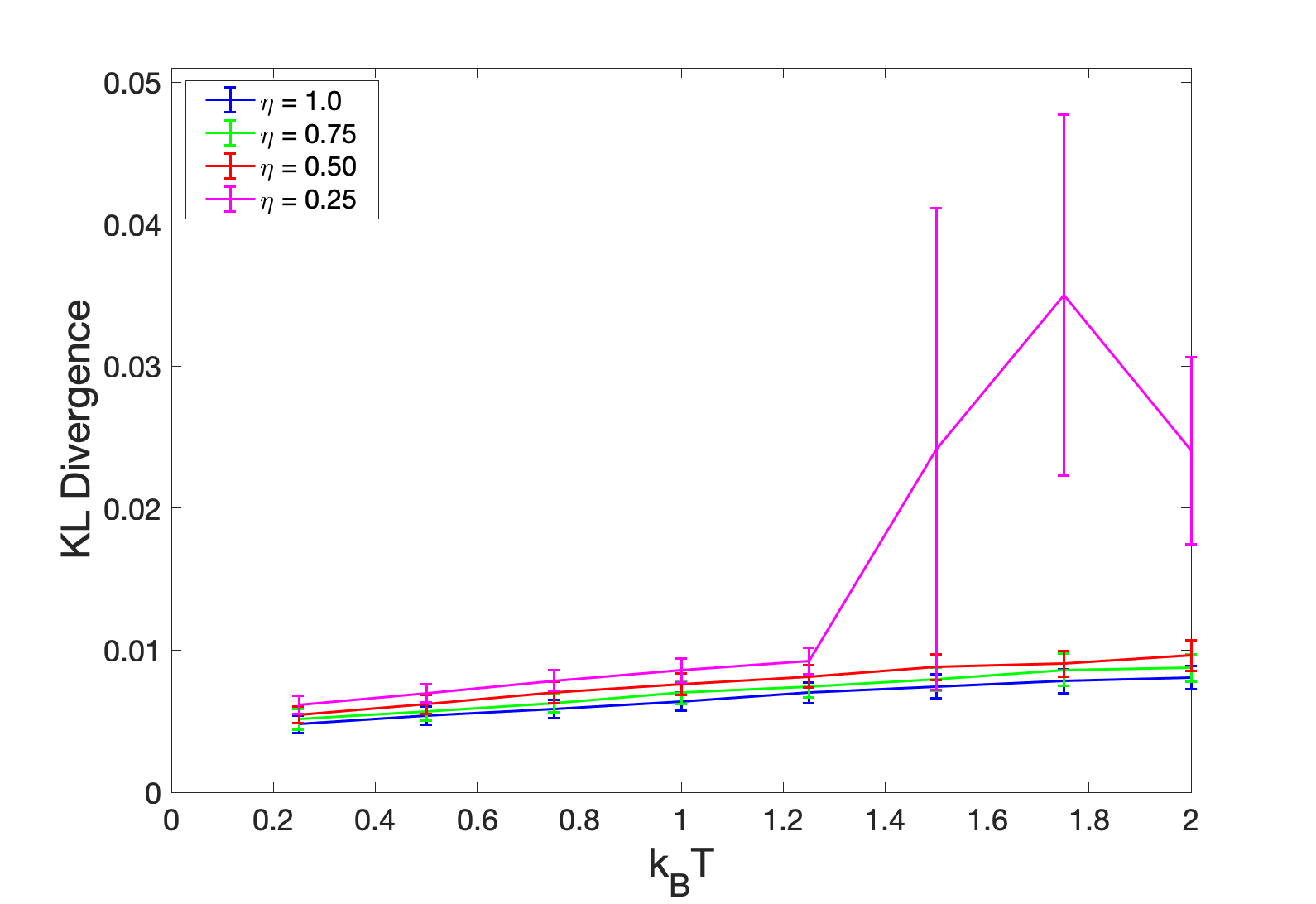} 
			\caption{\label{KLeta} Average values (with one standard deviation error bars) for KL divergences for the pdfs derived from the classical particle filter and the Wigner quasi-probability distributions for different values of the measurement efficiency $\eta$, other parameter values are given in the text.}
\end{figure}
Figure~\ref{KLeta} shows the equivalent results to Figure~\ref{KL} for $\gamma = 0.1$ and varying the measurement efficiency $\eta$. The results show a small increase in the accuracy of the estimated probability distribution as the efficiency is reduce, but there are no large increases until the efficiency drops to below $\eta = 0.5$, and then only for the larger temperature values ($k_B T > 1.25$). 

\subsection{Discussion of Results}\label{discussion}
The results presented are robust across a range of parameters and, in particular, across a range of temperatures. For brevity, we have not emphasised the feedback cooling aspect of the system, but one of the key factors in presenting the work in this paper is to provide a rationale for applying classical state estimation methods to assist in developing feedback control methods for a range of environmental conditions, including the quantum regime. Whilst the use of the SME to estimate the underlying quantum state would be preferred, it often proves impractical to use the SME for higher temperatures because the number of quantum states grows rapidly when the temperature is high enough to generate large excursions in energy. The ability to use classical state estimation methods across a wide range of temperatures to control the cooling mechanism, for example, would be beneficial. Some preliminary results using real-time Kalman filter based methods have already been demonstrated~\cite{Set2018}, and more sophisticated approaches have been proposed~\cite{Fer2019}. The bootstrap filter used for this work demonstrates the utility of the approach, and the relative simplicity of the filter should mean that the results presented in this paper could be improved using more sophisticated state estimation methods. Combining these methods with parameter estimation~\cite{Ral2017} and model selection~\cite{Ral2018} would also allow for better methods to test the limits of current physical theories of gravity~\cite{Ger2010,Arv2013} and modifications of quantum mechanics~\cite{Bas2003,Bas2013,Ber2015}

\section{Summary and Conclusions}\label{conclusions}
In this paper, we have used a conventional particle filter, a bootstrap filter, to generate probability density functions from continuous classical measurement records taken from a model quantum mechanical system in contact with a thermal environment. We have shown that the classical trajectories produced by the particle filter are good approximations to the quantum trajectories found by numerically integrating the stochastic master equation conditioned on the measurement record. We have also shown that the classical probability density functions are good approximations to the positive regions of the quantum mechanical Wigner quasi-probability distribution, and that the trajectories and the distributions are robust over a range of temperatures, nonlinearities, and measurement efficiencies. 

The ability to estimate quantum trajectories and aspects of quantum states using classical state estimation methods allows quantum feedback control to be extended to regions of parameter space for which the usual quantum state estimation methods are impractical. For the example discussed in this paper, a bead trapped in an optical trap, the ability to cool a material object down to temperatures close to the quantum regime is a fascinating prospect, and one that has recently been realised in these systems. The use of classical state estimation can help to explore this regime and potentially provide robust feedback control methods that do not require unrealistic computational resources.

\section*{Acknowledgment}
The authors would like to thank Dr. Mark Everitt, Dr. Russell Rundle, Dr. Ashley Setter, Dr. Muddassar Rashid, and Dr. Marko Toro\u{s} for useful discussions and advice during the preparation of this paper.



%

\end{document}